\documentclass[onecolumn,epjc3]{svjour3}
\usepackage{mathtext,bbm,amsmath,amsfonts,amssymb,indentfirst,syntonly,graphicx}
\smartqed  
\RequirePackage{graphicx}
\RequirePackage{subfigure}
\RequirePackage{xcolor}
\def\bc{\begin{center}}
\def\ec{\end{center}}
\def\be{\begin{eqnarray}}
\def\ee{\end{eqnarray}}

\journalname{Eur. Phys. J. C}

\begin{document}
\title{Constraining anisotropy of the universe from different groups of type-Ia supernovae}
\author{Zhe Chang \thanksref{addr1,addr2},
           Xin Li \thanksref{addr1,addr2},
           Hai-Nan Lin \thanksref{e1,addr1}, and
           Sai Wang \thanksref{addr1}
}
\thankstext{e1}{e-mail: linhn@ihep.ac.cn}
\institute{Institute of High Energy Physics, Chinese Academy of Sciences, 100049 Beijing, China\label{addr1}\and
           Theoretical Physics Center for Science Facilities, Chinese Academy of Sciences, 100049 Beijing, China\label{addr2}
}
\date{Received: date / Accepted: date}
\maketitle

\begin{abstract}
Recent released Planck data and other astronomical observations show that the universe may be anisotropic on large scales. Inspired by this, anisotropic cosmological models have been proposed. We note that the Finsler-Randers spacetime provides an appropriate framework for the anisotropic cosmology. By adding an arbitrary 1-form to the Friedmann-Robertson-Walker (FRW) line element, a privileged axis in the universe is picked out. The distance-redshift relation is modified to be direction-dependent. We wish that the anisotropic cosmological model may be tested crossly by independent observations. Type-Ia supernovae (SNe Ia) calibrated from four different light curve fitters are used to constrain possible anisotropy of the universe. The magnitudes of anisotropy are all between 2\% \--- 5\%, but the systematic uncertainty cannot be excluded. The directions of privileged axis seem to differ from each other. The statistical significance is not high enough to make a convincing conclusion. Nevertheless, the $1\sigma$ contours in the $(l,b)$ plane obtained from four groups of SNe Ia have an overlap, centering at $(l,b)\approx (170^{\circ},0^{\circ})$. Monte Carlo simulation shows that the anisotropy is unlikely to be caused by selection effect.
\end{abstract}

\section{Introduction}

The ${\rm \Lambda}$CDM model is well known as the standard model of modern cosmology. It is based on the fundamental assumption called cosmological principle, which states that the universe is homogeneous and isotropic at large scales. The general geometric structure of the universe can be described by the spherically symmetric Friedmann-Robertson-Walker (FRW) metric. The ${\rm \Lambda}$CDM model is well consistent with the seven-year WMAP observations \cite{Spergel:2007hy,Komatsu:2011fb} and the recent released Planck 2013 results \cite{Ade:2013ktc,Ade:2013nlj}. The tiny fluctuation of cosmic microwave background (CMB) radiation implies that the cosmological principle is an excellent approximation \cite{Smith:2009jr}. However, recent observations show that the universe may be deviated from isotropy. For example, the large-scale bulk flow \cite{Kashlinsky:2009ut,Watkins:2009hf,Lavaux:2010th}, the alignments of low multipoles in CMB angular power spectrum \cite{Lineweaver:1996xa,Tegmark:2003,Bielewicz:2004en,Copi:2010,Frommert:2010qw}, the large-scale alignments of the quasar polarization vectors \cite{Hutsemekers:2005iz,Hutsemekers:2008iv}, the spatial variation of fine-structure constant \cite{Dzuba:1999,Murphy:2001,Murphy:2003,King:2012}, the CMB hemispherical asymmetry observed by WMAP \cite{Bennett:2011,Bennett:2012zja} and Planck satellite \cite{Ade:2013nlj}.

Type-Ia supernovae (SNe Ia) are widely used as the standard candle to test possible anisotropy of the universe. Schwarz \& Weinhorst \cite{Schwarz:2007wf} used four groups of SNe Ia with redshift $z<0.2$ to test the isotropy of the Hubble diagram, and found a maximal hemispheric asymmetry towards a direction close to the equatorial poles. Using the hemisphere comparison method, Antoniou \& Perivolaropoulos \cite{Antoniou:2010} found that the highest expansion direction of the universe points towards $(l,b)=(309^{\circ}\,^{+23^{\circ}}_{-03^{\circ}}, 18^{\circ}\,^{+11^{\circ}}_{-10^{\circ}})$ in the Union2 compilation. The maximum anisotropy level is about $\Delta\Omega_M/\Omega_M\approx 0.43\pm 0.06$. Using the same method and dataset, Cai \& Tuo \cite{CaiTuo:2012} found that the maximum accelerating expansion direction points to $(l,b)=(314^{\circ}\,^{+20^{\circ}}_{-13^{\circ}}, 28^{\circ}\,^{+11^{\circ}}_{-33^{\circ}})$, and the maximum anisotropy is at the order of magnitude $\Delta q_0/q_0\approx 0.79^{+0.27}_{-0.28}$. Kalus et al. \cite{kalus:2013} tested the anisotropy of local universe using low-redshift ($z<0.2$) SNe Ia calibrated from four different light curve fitters. The highest expansion rate is found to be in the direction $(l, b)\approx (325^{\circ}, -19^{\circ})$\footnote{In the original paper of  Kalus et al., the authors used the convention that the galactic longitude $l$ is from $-180^{\circ}$ to $+180^{\circ}$. In order to make the comparison easier, we convert it to the range $l\in[0^{\circ},360^{\circ}]$.}, and the magnitude of Hubble anisotropy is about $\Delta H/H\approx 0.026$. Zhao et al. \cite{Zhao:2013yaa} studied the anisotropy of cosmic acceleration by dividing the Union2 dataset into 12 subsets according to their positions, and found a significant dipole effect in the $q_0$-maps. The direction of this dipole is nearly perpendicular to the CMB kinematic dipole. The redshift of SNe Ia is usually no more than 2. As a supplementary to SNe Ia, gamma-ray bursts (GRBs) are also used by some authors to test anisotropy of the universe \cite{Cai:2013lja}. However, there are many controversies in calibrating GRBs data. The systematic uncertainty of GRBs is much larger than that of SNe Ia.

In the theoretical aspect, some anisotropic cosmological models have been studied \cite{Mimoso:1993,Kumar:2011ui,Verma:2011,Singh:2012}. We note that the Finsler-Randers spacetime \cite{Randers:1941,Li:2010,Li:2012,Chang:2013} provides an appropriate framework for the anisotropic cosmology. The line element in Finsler-Randers spacetime can be described by the FRW line element with an extra 1-form \cite{Chang:2013}. This 1-form picks out a privileged axis so that the universe becomes axis-symmetric. The luminosity distance depends on not only the redshift, but also the direction. A direct fit to the Union2 dataset shows that the magnitude of anisotropy is about $D\approx 0.03\pm 0.03$, and the privileged axis points towards $(l,b)=(304^{\circ}\pm 43^{\circ},-27^{\circ}\pm 13^{\circ})$ in galactic coordinate system (GCS) \cite{Chang:2013}. This axis is close to the direction of highest expansion of the universe found by Kalus et al. \cite{kalus:2013}. Nevertheless, The statistical significance of this result is too low to be conclusive. The anisotropic magnitude is small enough such that the ${\rm \Lambda}$CDM model is still a good approximation.

We cannot make a convincing conclusion from only one dataset. Anisotropy may come from systematic uncertainty, as well as the intrinsic property of the universe. In this paper, we use different datasets published by literatures to constrain possible anisotropy of the universe. If the privileged axes derived from different datasets are close to each other, we can safely conclude that anisotropy is an intrinsic property of the universe. Otherwise, systematic uncertainty may dominate. The rest of the paper is arranged as follows: In section \ref{sec:model}, we briefly introduce the anisotropic cosmological model in the Finsler-Randers spacetime. In section \ref{sec:numerical}, SNe Ia data calibrated from four different light curve fitters are used to constrain the model parameters. We first fit the data of each group independently. Then, the intersection of four groups is picked out to fit the model. Finally, we re-analyze the data by restricting the redshift to $z<0.2$. In section \ref{sec:MCsimulation}, Monte Carlo simulation is performed to rule out the selection effect. Discussions and conclusions are given in section \ref{sec:conclusions}.

\section{Anisotropic cosmological model in the Finsler-Randers spacetime}\label{sec:model}

In the standard cosmological model, the spacetime is of Riemann type and the line element is described by FRW metric. In the Finsler-Randers spacetime, however, the line element can be written as the spatially flat FRW line element added by an extra 1-form \cite{Chang:2013}
\begin{equation}\label{eq:Randers}
  d\tau=\sqrt{dt^2-a^2(t)(dx^2+dy^2+dz^2)}+\tilde{b}_{\mu}({\mathbf x})dx^{\mu},
\end{equation}
where ${\mathbf x}=(t,x,y,z)$ is the four-dimensional spacetime coordinate. The 1-form $\tilde{b}_{\mu}({\mathbf x})dx^{\mu}$ on the right-hand-side of Eq.(\ref{eq:Randers}) picks out a privileged axis in the universe. For convenience, the direction of privileged axis is denoted by $\hat{\bf n}$. Here and after, we take the convention that a hat over a vector denotes the unit vector along that direction. Chang et al. \cite{Chang:2013} have showed that the anisotropy of Hubble diagram originates from spatial components of the privileged axis. Without loss of generality, one can choose a Cartesian coordinate system (CCS) such that the $z$-axis is exactly towards the privileged direction. Furthermore, assuming that the 1-form does not depend on the spacetime coordinates, then Eq.(\ref{eq:Randers}) simplifies to
\begin{equation}
  d\tau=\sqrt{dt^2-a^2(t)(dx^2+dy^2+dz^2)}+Ddz,
\end{equation}
where $D$ represents the anisotropic magnitude of the universe. And by the assumptions above, $D$ is a constant.

The redshift $z$ of a supernova relates to its direction $\hat{{\bf p}}$ and time $t$ as \cite{Chang:2013}
 \begin{equation}
   1+z(t,\hat{{\bf p}})=\frac{1}{a(t)}(1-D\cos\theta),
 \end{equation}
where $\theta$ is the angle between $\hat{\bf n}$ and $\hat{\bf p}$, and $a(t)$ is the scale factor of the universe. Following the similar procedure in the ${\rm \Lambda}$CDM model \cite{Weinberg:2008}, we can derive the distance-redshift relation in the Finsler-Randers spacetime. For a supernova locating at direction $\hat{{\bf p}}$ and redshift $z$ on the sky, the luminosity distance can be modified to be \cite{Chang:2013}
\begin{equation}\label{eq:distance-redshift}
  d_L(z,\hat{{\bf p}})=(1+z)\frac{c}{H_0}\int_0^z \frac{(1-D\cos\theta)^{-1}dz}{\sqrt{\Omega_M\left(\frac{1-D\cos\theta}{1+z}\right)^{-3}+\Omega_{\Lambda}}}.
\end{equation}
Note that when we derive Eq.(\ref{eq:distance-redshift}), we have neglected the terms of order $D^2$ or higher. From observational considerations, the universe should not be too deviated from isotropy. Thus, we have $\mid D\mid \ll 1$. When the anisotropy vanishes, i.e., $D=0$, Eq.(\ref{eq:distance-redshift}) returns back to that of ${\rm \Lambda}$CDM model. In practice, it is more convenient to transform the luminosity distance to the dimensionless distance modulus
\begin{equation}\label{eq:DistanceModulus}
  \mu(z,\hat{{\bf p}})\equiv5\log_{10}\left[\frac{d_L(z,\hat{{\bf p}})}{{\rm Mpc}}\right]+25.
\end{equation}

\section{Numerical constraints from SNe Ia}\label{sec:numerical}

\begin{figure}
\centering
  \includegraphics[width=12 cm]{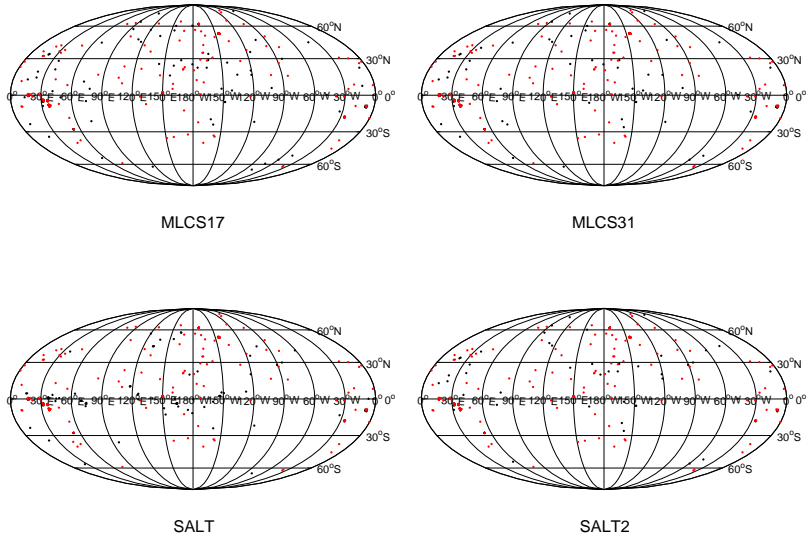}
  \caption{\small{Distribution of four groups of SNe Ia in the sky in ECS. Red dots denote the intersection of four groups. Only SNe Ia that satisfy the criteria ($A_V<0.5~\&~\Delta<0.7$ for MLCS17 and MLCS31, $-0.1<c<0.2$ for SALT and SALT2) are shown.}} \label{fig:distrubution}
\end{figure}

In this section, we use four groups of SNe Ia data published by literatures to constrain the model parameters. The data were originally published by Hicken et al. \cite{Hicken:2009}. They combined the CfA3 \cite{Hicken:2009aa} SNe Ia with the Union set \cite{Kowalski:2008} to form the so-called Constitution set. The distance-redshift relation was calibrated from four different light curve fitters: MLCS17, MLCS31, SALT and SALT2. There are 372, 366, 397 and 351 SNe Ia in each group, respectively. The intersection of four groups contains 258 SNe Ia. However, not all the data are suitable to be used to constrain the cosmological parameters. For example, MLCS overestimates the intrinsic luminosity of SNe Ia with $0.7<\Delta<1.2$, and MLCS31 overestimates $A_V$ while MLCS17 does not. Hicken et al. \cite{Hicken:2009} proposed the criteria that $A_V<0.5$ \& $\Delta<0.7$ for MLCS17 and MLCS31, while $-0.1<c<0.2$ for SALT and SALT2. Here, $A_V$ is the host galaxy extinction parameter, $\Delta$ is the shape/luminosity parameter, and $c$ is the color parameter. If we take the criteria into consideration, the number of SNe Ia in each group reduces to 324, 298, 309 and 265, respectively. The intersection of four groups consists of 183 SNe Ia. The highest redshift of SNe Ia in these four groups extends to 1.55, 1.39, 1.37 and 1.40, respectively. The distribution of SNe Ia in the sky of equatorial coordinate system (ECS) is plotted in Figure \ref{fig:distrubution}. SNe Ia that do not satisfy the criteria are neglected. Red dots correspond to the intersection of four groups. We can see that in each group, SNe Ia are approximately homogeneously distributed in the sky.

Kalus et al. \cite{kalus:2013} used low-redshift $(z < 0.2)$ SNe Ia in the Constitution set to test the isotropic expansion of the local universe. By comparing the best-fit Hubble diagrams in pairs of hemispheres, they found that the directions of highest expansion derived from four groups differ from each other. The axes referred from MLCS17 and SALT2 approximately point towards the same direction, i.e., $(l, b)=(308^{\circ},-19^{\circ})$ in galactic coordinate system (GCS). The direction obtained from SALT is $(l, b)=(206^{\circ},-32^{\circ})$, which is almost perpendicular to the directions obtained from MLCS17 and SLAT2. The anisotropic magnitudes derived by fitting the data of four groups seem to be close to each other, with an averaged Hubble anisotropy $\Delta H/H\approx 2.22\%$. This is consistent with the order of magnitude that can be expected by cosmic variance in the ${\rm \Lambda}$CDM universe. In order to compare with our results later, we list the results of Ref.\cite{kalus:2013} in Table \ref{tab:kalus}.

\begin{table}
\begin{center}
\caption{\small{The results of Ref.\cite{kalus:2013}. The first to fifth columns: the fitter, the total number of SNe Ia, the number of SNe Ia with $z<0.2$, the highest expansion direction in GCS, the magnitude of Hubble anisotropy.}}
\begin{tabular}[t]{ccccc}
\hline\hline
fitters & $N_{\rm total}$ & $N_{z<0.2}$ & $(l,b)$ & $\frac{H_N-H_S}{H_N+H_S}$\\
\hline
MLCS17 & 372 & 199 & $(308^{\circ},-19^{\circ})$  & 2.49\%\\
MLCS31 & 366 & 203 & $(342^{\circ},-20^{\circ})$  & 2.59\%\\
SALT   & 397 & 115 & $(206^{\circ},-32^{\circ})$  & 1.54\%\\
SALT2  & 351 & 183 & $(308^{\circ},-18^{\circ})$  & 2.28\%\\
\hline
\end{tabular}
\label{tab:kalus}
\end{center}
\end{table}

We try to fit the anisotropic cosmological model with the Constitution set. Since we are interested in large-scale anisotropy of the universe, we set no limit on the redshift. Before proceeding, some coordinate systems should be clarified. Hicken et al. \cite{Hicken:2009} have not provided the position of SNe Ia in their publication. We obtain the position from the list of SNe provided by the IAU Central Bureau for Astronomical Telegrams (CBAT)\footnote{http://www.cbat.eps.harvard.edu/lists/Supernovae.html}. Note that the position provided by the AIU CBAT is given in ECS. In order to compare with the results of Ref.\cite{kalus:2013}, we work directly in GCS. Thus, we should firstly convert the position of each SNe Ia from ECS to GSC. The detailed transformation between the two systems can be found in Ref.\cite{Peter:1981}. Corresponding to GCS, we define a CCS, whose origin locates at the center of GCS. The $z$-axis of CCS is towards north pole $(l,b)=(0^{\circ},90^{\circ})$, the $x$-axis is towards the point $(l,b)=(0^{\circ},0^{\circ})$, and the $xyz$ axes comprise the right-handed set. Thus, the orientation of a supernova with galactic coordinate $(l_i,b_i)$ can be rewritten in CCS as
\begin{equation}
  \hat{\bf p}=\cos(b_i)\cos(l_i)\hat{\bf i}+\cos(b_i)\sin(l_i)\hat{\bf j}+\sin(b_i)\hat{\bf k},
\end{equation}
where $\hat{\bf i}$, $\hat{\bf j}$ and $\hat{\bf k}$ are the unit vectors along the $x$, $y$ and $z$ axes, respectively. Furthermore, we suppose that the privileged axis $\hat{\bf n}$ points towards the direction ($l,b$) in GCS, which can also be rewritten in CCS as
\begin{equation}
  \hat{\bf n}=\cos(b)\cos(l)\hat{\bf i}+\cos(b)\sin(l)\hat{\bf j}+\sin(b)\hat{\bf k}.
\end{equation}
Thus, the cosine of the angle $\theta$ in Eq.(\ref{eq:distance-redshift}) is given by the inner product of $\hat{\bf p}$ and $\hat{\bf n}$, i.e., $\cos\theta=\hat{\bf p}\cdot\hat{\bf n}$.

\begin{figure}
\centering
  \includegraphics[width=12 cm]{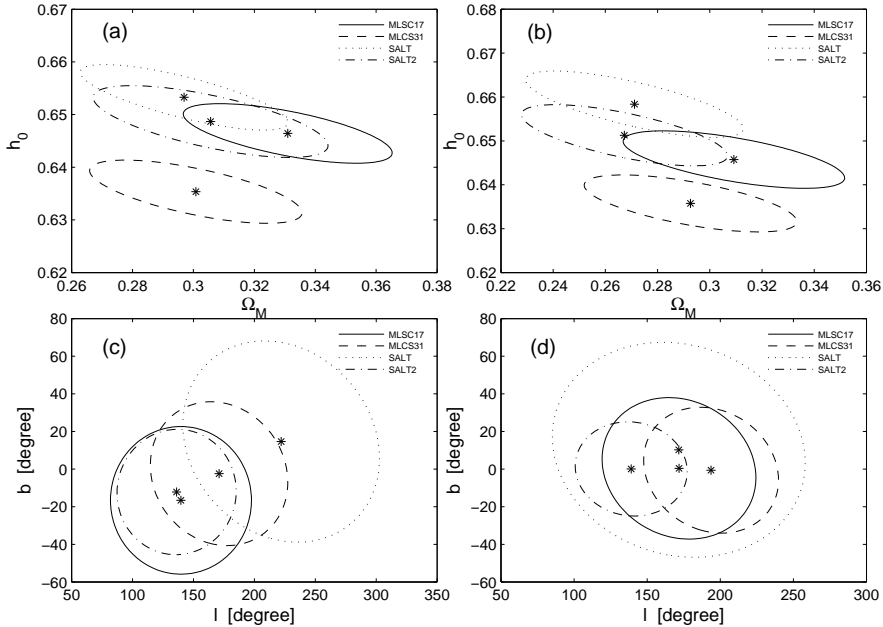}
  \caption{\small{The $1\sigma$ contours in the $(\Omega_M,h_0)$ and $(l,b)$ planes. Stars denote the central values. Panel (a): contours in the $(\Omega_M,h_0)$ plane derived from fitting four groups of SNe Ia independently. Panel (b): contours in the $(\Omega_M,h_0)$ plane derived from fitting to the Intersection set. Panel (c): contours in the $(l,b)$ plane derived from fitting four groups of SNe Ia independently. Panel (d): contours in the $(l,b)$ plane derived from fitting to the Intersection set.}}
  \label{fig:contour}
\end{figure}

The anisotropic cosmological model has five parameters in total: the matter component $\Omega_M$, the Hubble constant $H_0$, the anisotropic magnitude $D$, and the direction of privileged axis $(l,b)$. The least-$\chi^2$ method is used to find the model parameters. Define $\chi^2$ as
\begin{equation}
  \displaystyle\chi^2=\sum_{i=1}^N\left[\frac{\mu_{\rm th}^{(i)}-\mu_{\rm obs}^{(i)}}{\sigma_{\mu}^{(i)}}\right]^2,
\end{equation}
where $\mu_{\rm th}$ is the theoretical distance modulus calculated from Eqs.(\ref{eq:distance-redshift}) and (\ref{eq:DistanceModulus}), $\mu_{\rm obs}$ and $\sigma_{\mu}$ are respectively the observed distance modulus and its uncertainty, and $N$ is the total number of SNe. Firstly, we fit the data of each group independently. SNe Ia that do not satisfy the criteria are not used. As a zeroth order approximation, i.e., $D=0$, the anisotropic cosmological model reduces to the well-known ${\rm \Lambda}$CDM model. The best-fit parameters $\Omega_M$ and $h_0$ ($H_0=100h_0$ km s$^{-1}$ Mpc$^{-1}$) to ${\rm \Lambda}$CDM model are listed in the third and fourth columns of Table \ref{tab:choose-criteria}. The quoted uncertainties in the table are all of $1\sigma$. The $1\sigma$ contours in the $(\Omega_M,h_0)$ plane are plotted in panel (a) of Figure \ref{fig:contour}. Then we fix $(\Omega_M,h_0)$ to their central values and do a three-parameter fit. This gives constraints on the anisotropic magnitude $D$, and the privileged axis $(l,b)$. The results are listed in the fifth and sixth columns of Table \ref{tab:choose-criteria}. It should be noted that corresponding to each solution in Table \ref{tab:choose-criteria}, there is another equivalent solution, which $D$ changes its sign and the privileged axis changes to its opposite direction, ie., $D\longrightarrow -D$ and $(l,b)\longrightarrow (l-180^{\circ},-b)$. This can be easily seen from Eq.(\ref{eq:distance-redshift}), since $D$ is always multiplied by $\cos\theta$. For simplicity, we constrain $D$ to be positive. The $1\sigma$ contours in the $(l,b)$ plane are plotted in panel (c) of Figure \ref{fig:contour}.

\begin{table}
\begin{center}
\caption{\small{The cosmological parameters derived by fitting to the Constitution set. Data fitted with each light curve fitter are used to fit the model independently. SNe Ia that do not satisfy the criteria ($A_V<0.5~\&~\Delta<0.7$ for MLCS17 and MLCS31, $-0.1<c<0.2$ for SALT and SALT2) are not used. The errors are of $1\sigma$.}}
\begin{tabular}[t]{cccccc}
\hline\hline
fitters & $N$ & $\Omega_M$ & $H_0$ [km/s/Mpc] & $D$ & $(l,b)$\\
\hline
MLCS17 & 324 & $0.33\pm 0.02$ & $64.6\pm 0.4$ & $0.036\pm 0.022$ & $(140^{\circ}\pm 42^{\circ},-17^{\circ}\pm 28^{\circ})$\\
MLCS31 & 298 & $0.30\pm 0.03$ & $63.5\pm 0.4$ & $0.033\pm 0.024$ & $(171^{\circ}\pm 41^{\circ}, -2^{\circ}\pm 28^{\circ})$\\
SALT   & 309 & $0.30\pm 0.02$ & $65.3\pm 0.5$ & $0.022\pm 0.021$ & $(222^{\circ}\pm 58^{\circ}, 15^{\circ}\pm 39^{\circ})$\\
SALT2  & 265 & $0.31\pm 0.03$ & $64.9\pm 0.5$ & $0.046\pm 0.024$ & $(136^{\circ}\pm 35^{\circ},-12^{\circ}\pm 24^{\circ})$\\
\hline
\end{tabular}
\label{tab:choose-criteria}
\end{center}
\end{table}

From Table \ref{tab:choose-criteria}, we can see that four groups give similar constraints on parameters $\Omega_M$ and $H_0$, with average values $\Omega_M\approx 0.31$ and $H_0\approx 64.6$ km s$^{-1}$ Mpc$^{-1}$, respectively. The statistical uncertainty of $H_0$ is extremely small. This is in our expectation, since $H_0=65.0$ km s$^{-1}$ Mpc$^{-1}$ is prior adopted when calibrating the data. However, the constraint on anisotropy is not so strict. The anisotropic magnitudes constrained from four groups are close to each other, with an average value $D\approx 3.4\%$, although the statistical uncertainty is large. This result is approximately in accordance with that of Ref.\cite{kalus:2013}. The privileged axes constrained from MLCS17 and SALT2 almost point towards the same direction. The angle between these two axes is about $6^{\circ}$. The average direction is $(l,b)=(138^{\circ},-15^{\circ})$, or equivalently, ($318^{\circ},15^{\circ}$). This is consistent with the results of Ref.\cite{Antoniou:2010} and Ref.\cite{CaiTuo:2012}, but not of Ref.\cite{kalus:2013}. The other two groups, MLCS31 and SALT, give rather different directions. The direction constrained from SALT is almost perpendicular to that of MLCS17 and SALT2. Besides, the uncertainty of privileged axis is too large to make a convincing conclusion. Nevertheless, the $1\sigma$ contours in the $(l,b)$ plane obtained from four groups have an overlap, centering at $(l,b)\approx (170^{\circ},0^{\circ})$.

\begin{table}
\begin{center}
\caption{\small{The cosmological parameters derived by fitting to the Intersection set. There are 183 SNe Ia in each group. The errors are of $1\sigma$.}}
\begin{tabular}[t]{cccccc}
\hline\hline
fitters & $N$ & $\Omega_M$ & $H_0$ [km/s/Mpc] & $D$ & $(l,b)$\\
\hline
MLCS17 & 183 & $0.31\pm 0.03$ & $64.6\pm 0.5$ & $0.037\pm 0.028$ & $(172^{\circ}\pm 38^{\circ}, 0^{\circ}\pm 27^{\circ})$\\
MLCS31 & 183 & $0.29\pm 0.03$ & $63.6\pm 0.5$ & $0.040\pm 0.027$ & $(194^{\circ}\pm 33^{\circ}, 0^{\circ}\pm 24^{\circ})$\\
SALT   & 183 & $0.27\pm 0.03$ & $65.8\pm 0.6$ & $0.022\pm 0.026$ & $(172^{\circ}\pm 63^{\circ},10^{\circ}\pm 41^{\circ})$\\
SALT2  & 183 & $0.27\pm 0.03$ & $65.1\pm 0.5$ & $0.050\pm 0.023$ & $(139^{\circ}\pm 28^{\circ}, 0^{\circ}\pm 18^{\circ})$\\
\hline
\end{tabular}
\label{tab:intersection}
\end{center}
\end{table}

Although there is a significant overlap of SNe Ia in four groups, the difference between each group cannot be ignored. In order to check whether the different results come from different individuals in each group, we pick out the intersection of four groups, which we call Intersection set for convenience. This makes sure that all of four groups contain the same SNe Ia. The Intersection set consists of 183 SNe Ia. Following the similar procedure above, we firstly set $D=0$ and do a two-parameter fit to the ${\rm \Lambda}$CDM model. This gives constraint on parameters ($\Omega_M,h_0$). Then we fix ($\Omega_M,h_0$) to their central values and do a three-parameter fit to investigate possible anisotropy of the universe. The results are concluded in Table \ref{tab:intersection}. All of the errors quoted are of $1\sigma$. The $1\sigma$ contours in the $(\Omega_M,h_0)$ and $(l,b)$ planes are plotted in panel (b) and panel (d) of Figure \ref{fig:contour}, respectively. As the Constitution set, the Intersection set gives strict constraint on ${\rm \Lambda}$CDM model, with average values $\Omega_M=0.29$ and $H_0=64.8$ km s$^{-1}$ Mpc$^{-1}$. However, the constraint on anisotropy of the universe is not improved. The MLCS17 and SALT groups give a similar direction pointing towards $(l,b)=(172^{\circ},5^{\circ})$, or equivalently, $(352^{\circ},-5^{\circ})$. From panel (d) of Figure \ref{fig:contour}, we can see that the $1\sigma$ contours in the $(l,b)$ plane obtained from four fitters have an overlap, centering at $(l,b)\approx (170^{\circ},0^{\circ})$. Interestingly, this direction is consistent with that of Constitution set.

\begin{figure}
\centering
  \includegraphics[width=12 cm]{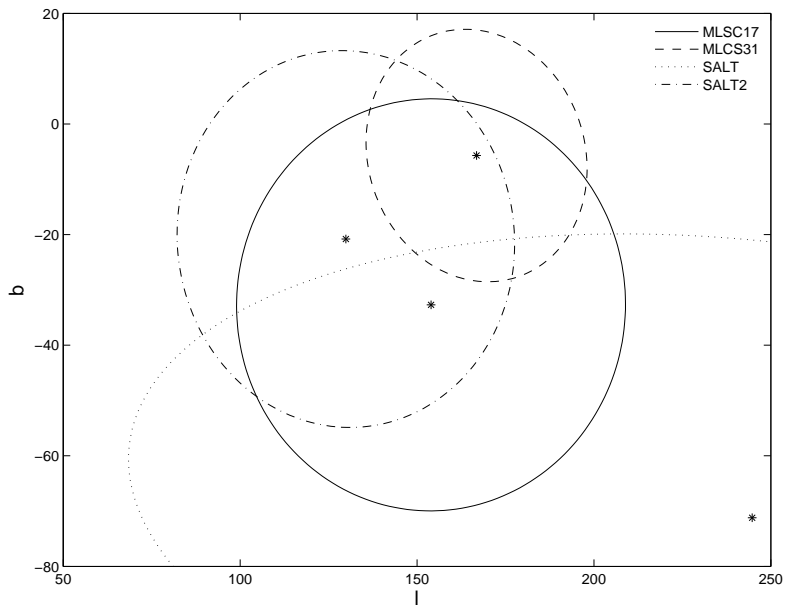}
  \caption{\small{The $1\sigma$ contours in the $(l,b)$ plane derived by fitting to the low-redshift $(z<0.2)$ SNe Ia. Stars denote the central values.}}
  \label{fig:contour2}
\end{figure}

It is not surprising that the privileged axis find here differs from that of Kauls et al. \cite{kalus:2013},  since we probe a different range of redshift. In Ref.\cite{kalus:2013}, the authors only analyzed the low-redshift $(z<0.2)$ SNe Ia, while in this paper we give no restriction on the redshift. To make a direct comparison, we re-analyze the data by restricting the redshift to $z<0.2$. As is done above, we first set $D=0$ and fit the data to the ${\rm \Lambda}$CDM model. Unfortunately, we find that the dataset cannot give strict constraint on  ${\rm \Lambda}$CDM model. The best-fit value of $\Omega_M$ is far away from the widely accepted value (i.e., $\Omega_M \approx 0.3$). For the MLCS17 group, $\Omega_M$ is as small as 0.03. The worst thing is that the MLCS31 group gives a negative $\Omega_M$.  This is not amazing since the structure of the local universe may differ from the large-scale structure. In order to avoid such an unreasonableness, we fix $\Omega_M$ and $h_0$ to their average value given in Table \ref{tab:choose-criteria}, i.e., $(\Omega_M,h_0)=(0.31, 0.646)$. Then we fit the data to the anisotropic cosmological model to derive the anisotropic magnitude $D$ and the direction of privileged axis $(l,b)$. The best-fit parameters are listed in Table \ref{tab:low-reshift}, and the $1\sigma$ contours in the $(l,b)$ plane are plotted in Figure \ref{fig:contour2}. It can be seen that three out of the four groups (except for SALT) give the similar directions, with an average direction $(l,b)=(150^{\circ},-20^{\circ})$, or equivalently, $(l,b)=(330^{\circ},20^{\circ})$. This direction is not far away from the direction found in Ref.\cite{kalus:2013}, i.e., $(l,b)=(325^{\circ},-19^{\circ})$. Interestingly, the longitude of these two directions is very close to each other, while the latitude seems to have an opposite sign. One reason for the discrepancy is the different criteria we adopted to choose the data. The rest group, SALT, gives a rather different direction. Besides, the uncertainty of SALT is much larger than that of the other three.

\begin{table}
\begin{center}
\caption{\small{The anisotropic magnitude and privileged axis derived by fitting to the low-redshift ($z<0.2$) SNe Ia. The errors are of $1\sigma$. We set $(\Omega_M,h_0)=(0.310,0.646)$ in the calculation. Note that the number of low-redshift SNe in our paper differs from that of reference\cite{kalus:2013}, because we adopt different criteria to choose the data.}}
\begin{tabular}[t]{cccc}
\hline\hline
fitters & $N(z<0.2)$ & $D$ & $(l,b)$\\
\hline
MLCS17 & 165 & $0.042\pm 0.020$ & $(154^{\circ}\pm 40^{\circ},-33^{\circ}\pm 27^{\circ})$\\
MLCS31 & 150 & $0.069\pm 0.027$ & $(167^{\circ}\pm 23^{\circ}, -6^{\circ}\pm 17^{\circ})$\\
SALT   & 124 & $0.035\pm 0.020$ & $(245^{\circ}\pm 128^{\circ}, -71^{\circ}\pm 37^{\circ})$\\
SALT2  & 144 & $0.048\pm 0.022$ & $(130^{\circ}\pm 35^{\circ},-21^{\circ}\pm 25^{\circ})$\\
\hline
\end{tabular}
\label{tab:low-reshift}
\end{center}
\end{table}

\section{Monte Carlo simulation}\label{sec:MCsimulation}

A noticeable feature is that, the direction we found here approximately locates in the galactic plane. This may be caused by the selection effect, i.e., the lack of SNe detected towards the center of Milky Way may introduce such an asymmetry. In order to check whether the anisotropy is due to the selection effect, Monte Carlo simulation should be carried out. A convenient way to do so is to randomly scramble the data points in redshift and distance modulus, but to keep their positions fixed on the sky, and then reanalyze the scrambled data. Repeating the procedure $N$ times, we can derive $N$ directions, as well as $N$ magnitudes. If a significant fraction of cases of such a Monte Carlo simulation show the same direction, it is high likely that the anisotropy is a selection artefact. On the contrary, if the directions derived from the simulations are uniformly distributed in the sky, the selection effect can be safely ruled out.

\begin{figure}
\centering
  \includegraphics[width=12 cm]{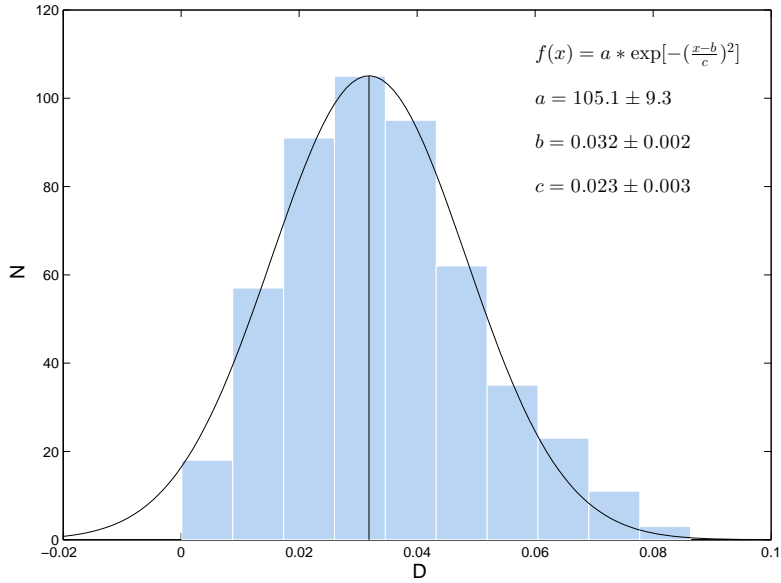}
  \caption{\small{The anisotropic magnitude $D$ in the 500 simulations follows the Gauss distribution, with an average value $\bar{D}=0.032$, and standard variance $\sigma$=0.016. The black curve is the best fit to the histogram. The black vertical line is the center of the Gauss distribution.}}
  \label{fig:D_distribution}
\end{figure}

\begin{figure}
\centering
  \includegraphics[width=12 cm]{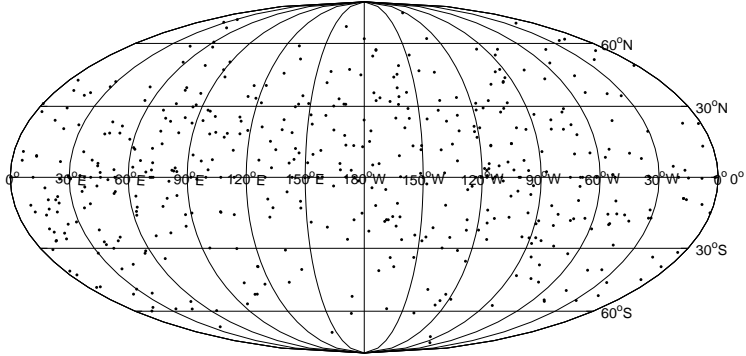}
  \caption{\small{The privileged axes in the 500 simulations seem to be uniformly distributed in the sky of GCS.}}
  \label{fig:lb_distribution}
\end{figure}

We take the MLCS17 group as an example. After selection, the MLCS17 group contains 324 SNe. As was showed in the last section, a direct fit to the ${\rm \Lambda}$CDM model gives the best-fit parameters $\Omega_M=0.33$ and $H_0=64.6$ km s$^{-1}$ Mpc$^{-1}$ (see Table \ref{tab:choose-criteria}). We fix ($\Omega_M,H_0$) to these values in Monte Carlo simulation. In each simulation, we fit the scrambled data to the anisotropic cosmological model to get the anisotropic magnitude $D$ and the preferred direction $(l,b)$. We repeat the simulation 500 times and obtain 500 groups of best-fit parameters $(D_i,l_i,b_i)$, where $i=1,2,\cdots,500$. We divide $D_i$ into 10 uniformly-spaced bins and plot the histogram in Figure \ref{fig:D_distribution}. As can be seen, the histogram can be well fitted by the Gauss function
\begin{equation}
f(D)=a\exp\left[-\left(\frac{D-b}{c}\right)^2\right].
\end{equation}
The best-fit values and their $1\sigma$ uncertainties are
\begin{equation}
  a=105.1\pm 9.3,~~b=0.032\pm 0.002,~~c=0.023\pm 0.003.
\end{equation}
The averaged magnitude of anisotropy, $\bar{D}=b=0.032$, is consistent with the results of the last section. The standard variance is $\sigma=c/\sqrt{2}=0.016$. The directions derived in the 500 simulations are plotted in Figure \ref{fig:lb_distribution}. It seems that the directions are uniformly distributed in the sky. Therefore, it is unlikely that the anisotropy is a result of selection effect.

\section{Discussions and conclusions}\label{sec:conclusions}

Recent observations on large-scale structure of the universe imply that the cosmos may be anisotropic. As an intrinsically anisotropic geometry, the Finsler geometry provides us an ideal framework to describe the anisotropic universe. An anisotropic cosmological model was proposed in the background of the Finsler-Randers spacetime. An arbitrary 1-form adding to the FRW line element picks out a privileged axis in the universe, such that the universe becomes axis-symmetric. Giving some assumptions to the 1-form, the distance-luminosity relation was modified to be direction-dependent. SNe Ia data calibrated from four different light curve fitters were used to test possible anisotropy of the universe. The anisotropy constrained from four groups is found to be at the same order of magnitude, while the directions of privileged axis differ from each other. The statistical uncertainty is too large to make a convincing conclusion. Picking out the intersection of four groups does not significantly improve the results. Interestingly, the $1\sigma$ contours in the $(l,b)$ plane obtained from four groups overlap with each other, centering at $(l,b)\approx (170^{\circ},0^{\circ})$. This direction is approximately in the galactic plane. Monte Carlo simulation excludes the selection effect.

\begin{acknowledgements}
We are grateful to Zhao D. for useful discussions. This work has been funded by the National Natural Science Fund of China under Grant No. 11075166, No. 11305181, and No. 11375203.
\end{acknowledgements}


\begin{thebibliography}{}
\bibitem{Spergel:2007hy}D. N. Spergel, et al., [WMAP Collaboration], Astrophys. J. Suppl. {\bf 170}, 377 (2007)
\bibitem{Komatsu:2011fb}E. Komatsu, et al., [WMAP Collaboration], Astrophys. J. Suppl. {\bf 192}, 18 (2011)
\bibitem{Ade:2013ktc}P. A. R. Ade, et al.,  [Planck Collaboration], arXiv:1303.5062 (2013).
\bibitem{Ade:2013nlj}P. A. R. Ade, et al., [Planck Collaboration], arXiv: 1303.5083 (2013)
\bibitem{Smith:2009jr}K. M. Smith, L. Senatore \& M. Zaldarriaga, J. Cosmol. Astropart. Phys. {\bf 0909}, 006 (2009)
\bibitem{Kashlinsky:2009ut}A. Kashlinsky, F. Atrio-Barandela, D. Kocevski \& H. Ebeling, Astrophys.\ J.\ {\bf 686}, L49 (2008)
\bibitem{Watkins:2009hf}R. Watkins, H. A. Feldman \& M. J. Hudson, Mon.\ Not.\ Roy.\ Astron.\ Soc.\ {\bf 392}, 743 (2009)
\bibitem{Lavaux:2010th}G. Lavaux, R. B. Tully, R. Mohayaee \& S. Colombi, Astrophys.\ J.\ 709, 483 (2010)
\bibitem{Lineweaver:1996xa}C. H. Lineweaver, L. Tenorio, G. F. Smoot, P. Keegstra, A. J. Banday \& P. Lubin, Astrophys.\ J.\ {\bf 470}, 38 (1996)
\bibitem{Tegmark:2003}M. Tegmark, A. de Oliveira-Costa \& A. Hamilton, Phys. Rev. D {\bf 68}, 123523 (2003)
\bibitem{Bielewicz:2004en}P. Bielewicz, K. M. Gorski \& A. J. Banday, Mon.\ Not.\ Roy.\ Astron.\ Soc.\ {\bf 355}, 1283 (2004)
\bibitem{Copi:2010}C. J. Copi, D. Huterer, D. J. Schwarz \& G. D. Starkman, Advances in Astro. {\bf 2010}, 17 (2010)
\bibitem{Frommert:2010qw}M. Frommert \& T. A. En{\ss}lin, Mon.\ Not.\ Roy.\ Astron.\ Soc.\ {\bf 403}, 1739 (2010)
\bibitem{Hutsemekers:2005iz}D. Hutsemekers, R. Cabanac, H. Lamy \& D. Sluse, Astron.\ Astrophys.\ {\bf 441}, 915 (2005)
\bibitem{Hutsemekers:2008iv}D. Hutsemekers, A. Payez, R. Cabanac, H. Lamy, D. Sluse, B. Borguet \& J. R. Cudell, arXiv: 0809.3088 (2008)
\bibitem{Dzuba:1999}V. A. Dzuba, V. V. Flambaum \& J. K. Webb, Phys. Rev. Lett. {\bf 82}, 888 (1999)
\bibitem{Murphy:2001}M. T. Murphy, et al., Mon. Not. R. Astron. Soc. {\bf 327}, 1208 (2001)
\bibitem{Murphy:2003}M. T. Murphy, J. K. Webb \& V. V. Flambaum, Mon. Not. R. Astron. Soc. {\bf 345}, 609 (2003)
\bibitem{King:2012}J. A. King, et al., Mon. Not. R. Astron. Soc. {\bf 422}, 3370 (2012)
\bibitem{Bennett:2011}C. L. Bennett, et al., [WMAP Collaboration], Astrophys. J. Suppl. {\bf 192}, 16 (2011)
\bibitem{Bennett:2012zja}C. L. Bennett, et al., [WMAP Collaboration], Astrophys.\ J.\ Suppl.\  {\bf 208}, 20 (2013)
\bibitem{Schwarz:2007wf}D. J. Schwarz \& B. Weinhorst, Astron.\ Astrophys.\ {\bf 474}, 717 (2007)
\bibitem{Antoniou:2010}I. Antoniou \& L. Perivolaropoulos, J. Cosmol. Astropart. Phys. {\bf 1012}, 012 (2010)
\bibitem{CaiTuo:2012}R.-G. Cai \& Z.-L. Tuo, J. Cosmol. Astropart. Phys. {\bf 1202}, 004 (2012)
\bibitem{kalus:2013}B. Kalus, et al., Astron. Astrophys. {\bf 553}, A56 (2013)
\bibitem{Zhao:2013yaa}W. Zhao, P. X. Wu \& Y. Zhang, Int.\ J.\ Mod.\ Phys.\ D  {\bf 22}, 1350060 (2013)
\bibitem{Cai:2013lja}R.-G. Cai, Y.-Z. Ma, B. Tang \& Z.-L Tuo., Phys. Rev. D {\bf 87}, 123522 (2013)
\bibitem{Mimoso:1993}J. P. Mimoso \& P. Crawford, Class. Quantum Grav. {\bf 10}, 31 (1993)
\bibitem{Kumar:2011ui}S. Kumar, Mod.\ Phys.\ Lett.\ A {\bf 26}, 779 (2011)
\bibitem{Verma:2011}M. K. Verma, M. Zeyauddin \& S. Ram, Rom. J. Phys. {\bf 56}, 616 (2011)
\bibitem{Singh:2012}M. K. Singh, M. K. Verma \& S. Ram, Adv. Studies Theor. Phys. {\bf 6}, 117 (2012)
\bibitem{Randers:1941}G. Randers, Phys. Rev. {\bf 59}, 195 (1941)
\bibitem{Li:2010}X. Li \& Z. Chang, Phys. Lett. B {\bf 692}, 1 (2010)
\bibitem{Li:2012}X. Li \& Z. Chang, Differ. Geom. Appl. {\bf 30}, 737 (2012)
\bibitem{Chang:2013}Z. Chang, M.-H. Li, X. Li \& S. Wang, Eur. Phys. J. C {\bf 73}, 2459 (2013)
\bibitem{Weinberg:2008}S. Weinberg, {\it Cosmology}, Oxford University Press, New York (2008)
\bibitem{Hicken:2009}M. Hicken, et al., Astrophys. J. {\bf 700}, 1097 (2009)
\bibitem{Hicken:2009aa}M. Hicken, et al., Astrophys. J. {\bf 700}, 331 (2009)
\bibitem{Kowalski:2008}M. Kowalski, et al., Astrophys. J. {\bf 686}, 749 (2008)
\bibitem{Peter:1981}D. S. Peter, {\it Practical Astronomy with your Calculator}, Cambridge University Press, second edition (1981)

\end{thebibliography}

\end{document}